\documentclass[%
 reprint,
 amsmath,amssymb,
 aps,
]{revtex4-2}

\usepackage{graphicx}
\usepackage{dcolumn}
\usepackage{bm}
\usepackage[dvipsnames]{xcolor}

\newcommand{\Pin}{P_{\rm in}(t)}
\newcommand{\Nnew}{N_{\rm new}^{(\phi )}(t)}
\newcommand{\meanN}{N_{L}}
\newcommand{\Dshort}{D}

\usepackage{xr}
\makeatletter

\newcommand*{\addFileDependency}[1]{
\typeout{(#1)}
\@addtofilelist{#1}
\IfFileExists{#1}{}{\typeout{No file #1.}}
}\makeatother
\newcommand*{\myexternaldocument}[1]{%
\externaldocument{#1}%
\addFileDependency{#1.tex}%
\addFileDependency{#1.aux}%
}

\myexternaldocument{Supplementary}

\begin{document}
\title{Collective Hard Core Interactions Leave \\ Multiscale Signatures in Number Fluctuation Spectra}

\author{Eleanor K. R. Mackay}
\affiliation{%
Physical and Theoretical Chemistry Laboratory, South Parks Rd, Oxford, OX1 3QZ UK}
\author{Anna Drummond Young}
\affiliation{%
Physical and Theoretical Chemistry Laboratory, South Parks Rd, Oxford, OX1 3QZ UK}

\author{Adam Carter}%
\affiliation{%
CNRS, Sorbonne Université, Physicochimie des Electrolytes et Nanosystèmes Interfaciaux, F-75005 Paris, France
}

\author{Sophie Marbach}%
\affiliation{%
CNRS, Sorbonne Université, Physicochimie des Electrolytes et Nanosystèmes Interfaciaux, F-75005 Paris, France
}

\author{Alice L. Thorneywork}%
 \email{alice.thorneywork@chem.ox.ac.uk}
\affiliation{%
Physical and Theoretical Chemistry Laboratory, South Parks Rd, Oxford, OX1 3QZ UK}

\begin{abstract}
A full understanding of transport in dense, interacting suspensions requires analysis frameworks sensitive to self and collective dynamics across all relevant spatial and temporal scales. Here we tackle this problem with the power spectral density of particle number fluctuations (N-PSD). By combining colloidal experiments and theory we show that the N-PSD naturally probes behaviour across multiple important dynamic regimes and we fully uncover the mechanistic origins of characteristic spectral scalings and timescales. In particular, we demonstrate that while high-frequency scalings link to self-diffusion, low-frequency scalings sensitively capture long-lived correlations and collective dynamics. In this regime, interactions lead to non-trivial spectral signatures, governed by pairwise particle exchange at small length scales and collective rearrangements over large scales.  Our findings thus provide important insight into the effect of interactions on microscopic dynamics and fluctuation phenomena and establish a powerful new tool with which to probe dynamics in complex systems. 
\end{abstract}

\maketitle

In dense molecular and mesoscopic systems, interparticle interactions induce surprising dynamic phenomena, including anomalous sub-diffusion in crowded environments and Fickian non-Gaussian diffusion with heterogeneous particle mobilities \cite{Minton2006,Szymanski2009,Waigh2023,Weiss2004,Golding2006,Hofling2013,Saxton2007,wang2012brownian,Wang2009}. Here, colloidal microscopy has played a pivotal role \cite{Skinner2013,Sentjabrskaja2016,Guan2014,Pastore2021,pastore2022multiscale,Rusciano2022,Alexandre2023,Henderson2002}, enabling resolution of dynamics for single particles with controllable interactions. Many studies, however, still focus on a single-particle observable, the particle trajectory, and combine this with a single correlation function, the mean-squared displacement (MSD)~\cite{rose2020particle}. While sufficient in dilute suspensions where a single self-diffusion coefficient completely describes the dynamics, this approach reveals only limited aspects of behaviour in dense suspensions or complex environments\cite{Minton2006,Szymanski2009,Waigh2023,Weiss2004,Golding2006,Hofling2013,Saxton2007, Carter2025, manzo2015review,wang2012brownian}. In these cases, a complete description requires alternative analysis frameworks that are explicitly sensitive to interactions and capable of connecting self and collective behaviours across all relevant length and time scales.

Two common directions are either (i) to replace trajectories with observables derived from positions that access complex or multi-particle dynamics or (ii) to shift the correlation function from time to frequency domain to better resolve multiple time scales. However, current combinations of observable and correlation function still give an incomplete picture. For example, as observables, fluctuations in (fluorescence) intensity \cite{Digman2011,lu2012characterizing} or particle number \cite{Mackay2024,Carter2025} within a finite volume are inherently sensitive to many-body phenomena. Accessing these collective behaviours using time-correlation functions, however, requires accurate resolution of their long-time decay which is very challenging~\cite{Carter2025}. An alternative analysis sensitive to collective effects is the intermediate scattering function, the spatial Fourier transform of density correlations~\cite{dhont1996introduction,lattuada2025hitchhiker}. Here, the aperiodicity of experimental images introduces artefacts at large length scales~\cite{Carter2025,moisan2011periodic,giavazzi2017image,zuccolotto2024improving}. Analysis of trajectories via the power spectral density (PSD) -- the temporal Fourier transform of the time-autocorrelation function -- has recently proven powerful in soft matter, revealing anomalous diffusion and hydrodynamic correlations ~\cite{Franosch2011,Tolic2004,Krapf2018,Krapf2019,Sposini2019,Fox2021,Sposini2022,metzler2019brownian}.
Yet trajectories naturally probe self rather than collective phenomena. Moreover, even in settings where PSD-based analysis is long established (\textit{e.g.} current fluctuations in nanofluidic or solid-state systems~\cite{Burgess1965,Voss1976}) spectral signatures are difficult to interpret unambiguously. This leaves PSDs as a probe of collective dynamics largely unexplored.

\begin{figure*}
    \centering
    \includegraphics[width=\linewidth]{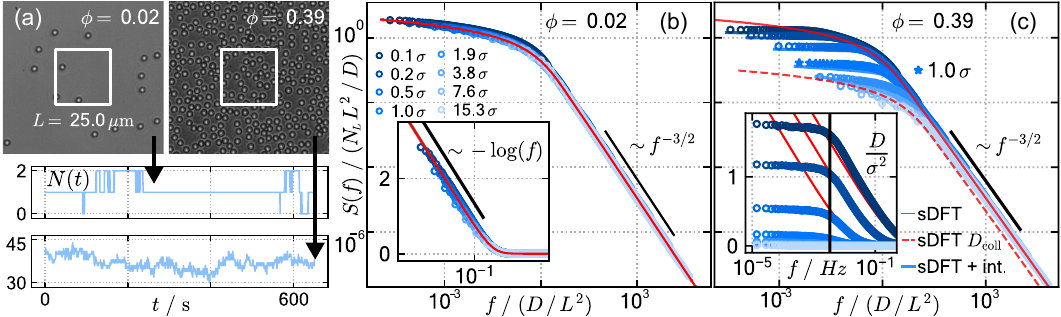}
    \caption{\textbf{Interactions modify low-frequency signatures in the power spectral density of particle number fluctuations} (a) Sections of microscopy images at low ($\phi=0.02$) and intermediate ($\phi=0.39$) packing fractions, with representative experimental particle counts $N(t)$ for a square box of size $L=25~\mu$m. Experimental power spectral densities of particle number $N(t)$ at (b) $\phi=0.02$ and (c) $\phi=0.39$ with rescaled amplitude and frequency. Points show experimental data for logarithmically spaced box sizes from $L = 0.3$~$\mu$m(dark) to $46.1~\mu$m (light), with values given in the legend. (Insets) Same data on linear-logarithmic axes, with unscaled frequency axis in (c). Red solid lines show theoretical results from sDFT (without interactions) using $\Dshort$ as input. In (c), blue lines are from sDFT with interactions \cite{SuppMat} and the dashed red line shows the large box size limit using the collective diffusion coefficient $D_{\rm coll}=D(1+\phi)/(1-\phi)^3$ \cite{Carter2025}.   }   
    \label{fig:fig1}
\end{figure*}

Here, we use theory and colloidal experiments to show how interactions shape the power spectral density of particle number fluctuations (N-PSD), a combination of observable plus correlation function that naturally captures self and collective dynamics across all relevant scales. In dilute suspensions, we resolve two diffusion-mediated spectral regimes, including persistent low-frequency correlations as $-\log(f)$. We link this non-trivial scaling to particle returns at long timescales \cite{Burgess1965,Voss1976}. At higher densities, the N-PSD exhibits new characteristic time scales, length scale dependence, and rapid decorrelation in specific regimes, all captured by analytical theory. This motivates novel analysis of particle trajectories, uncovering a subtle interplay between particle returns and neighbour arrivals, mediated by particle-particle interactions. We thereby establish the N-PSD as a powerful tool to uncover fundamental links between microscopic interactions and particle dynamics from microscopy

Our quasi-2D colloidal system consists of melamine particles (Microparticles GmbH) in 20/80\% v/v ethanol/water solutions confined by gravity to form a monolayer above the base of a quartz glass sample cell \cite{Thorneywork2014}. Particles interact as hard spheres with diameter $\sigma\sim3~\mu$m  and are considered at packing fractions, $\phi = N_0 \pi \sigma^2/(4A_0) $, from 0.02 to 0.66 where $A_0$ is the imaged area and $N_0$ the average number of particles in the image. The system is imaged using a custom-built bright-field inverted microscope at two frames per second for up to 24 hours and particle trajectories are obtained using standard routines \cite{Crocker1996,Trackpy2023}. 
Theoretical predictions are obtained within the framework of stochastic density functional theory (sDFT) \cite{dean1996langevin,Mackay2024} (see SI for further details \cite{SuppMat}). Brownian Dynamics simulations are performed using standard Euler-Maryama integration of overdamped Langevin equations of non-interacting particles as in Ref. \cite{DrummondYoung2025}.

To probe transport within the monolayer, we count the number of particles, $N(t)$, whose centres occupy regions of size $L_1$ × $L_2$ as a function of time (Fig.~\ref{fig:fig1}(a)) \cite{Mackay2024}. By counting based on the positions of particle centres, we can obtain signals $N(t)$ even in boxes much smaller than the particle size.
The power spectral density (N-PSD), $S(f)$, is then computed as
\begin{equation}
	\label{Eq:fourierdef}
S( f ) = \left \langle \frac{1}{T} \left|\int_0^T  N(t) e^{-i 2\pi f t} \, \mathrm{d}t \,\right|^2 \right \rangle,
\end{equation}
measured over sufficiently long times $T$, and where $\langle \cdot \rangle$ indicates an average over multiple windows in time \cite{Press1992} and multiple overlapping boxes ($N_{\rm boxes} \sim 10^3-10^4$ )\cite{Carter2025}. Anti-aliasing filters are constructed as in \cite{Kirchner2005}.

Fig.~\ref{fig:fig1} shows spectra for a range of box sizes ($L_1=L_2=L$) at two packing fractions: a very dilute system ($\phi=0.02$) in (b) and an intermediate density ($\phi=0.39$) in (c). At $\phi=0.02$ spectra are in excellent agreement with analytical theory for non-interacting systems \cite{Mackay2024}. They show perfect collapse when frequency is rescaled by $f_D=\Dshort/L^2$ and $S(f)$ by a characteristic power density $\meanN/f_D$, where $\Dshort$ is the short-time self-diffusion coefficient (see \cite{SuppMat}) and $\meanN \equiv \langle N \rangle$ the mean number of particles in the box. This indicates that there is a single characteristic time scale in the system: the time for particles to diffuse the box width, $\tau_D=L^2/\Dshort$.
Rescaled data show two scaling regimes, each resolved over several orders of magnitude: a high-frequency regime as $f^{-3/2}$ for $f>f_D$ and a low-frequency regime as $-{\rm log}(f)$ for $f<f_D$, shown on linear-logarithmic axes in Fig.~\ref{fig:fig1}(b), inset. 
Remarkably, the low-frequency noise highlights that correlations persist in this system over very long time scales, here up to $24\,\mathrm{hrs}$. 

For higher packing fractions, the high-frequency regime continues to conform to a $f^{-3/2}$ scaling and can be described with sDFT without explicit interparticle interactions using the experimental short-time self-diffusion coefficient $\Dshort$ as input. This accounts for the reduced particle mobility arising from hydrodynamic interactions \cite{Dhont1996,Mackay2024}. For lower frequencies, however, spectra show striking differences to those at low $\phi$, including a non-trivial dependence of scalings on box size. 
For large boxes, we observe the same $-\log(f)$ scaling as at low densities, but with reduced amplitude and higher corner frequency. 
We find this characteristic time scale corresponds to the $\phi$-dependent collective diffusion coefficient $D_{\rm coll}$~\cite{Carter2025} (red dashed line in Fig.~\ref{fig:fig1}(c)). Density fluctuations at large length scales thus relax diffusively at long times but do so faster than in the non-interacting case.

More interestingly, for small box sizes $L \lesssim \sigma$ we observe an apparent plateau in the spectra at low frequency (Fig.~\ref{fig:fig1}(c, inset)) -- a somewhat surprising observation given that we might expect more significant correlation in the interacting system. Fig. 1(c, inset) also shows that for small box sizes, the corner frequency is now independent of box size and scales as $f_\sigma=D/\sigma^2$, \textit{i.e.} the relevant timescale is that for a particle to diffuse over its diameter, $\sigma$. Since $D$ can be obtained independently from the high-frequency regime \cite{Mackay2024}, this corner frequency gives a measure of the particle size.

To shed light on this behaviour, we investigate how different particle transport processes contribute to the N-PSD. The PSD links to the time autocorrelation function via $S(f)=\int\langle N(t)N(0)\rangle e^{-i2\pi ft}\,dt$  \cite{Bendat2011}. As such, the two scaling regimes in the PSD reflect two distinct time scalings in the correlation function $\langle N(t)N(0)\rangle$. This can be recast in terms of individual particle contributions as $N(t)=\sum_in_i(t)$, where $n_i(t)=\{1,0\}$ when particle $i$ is inside or outside the box respectively. Expanding the correlation function explicitly, we can split correlations into those arising from the same or different particles occupying the box at later time $t$,
\begin{equation}
    \langle N(t)N(0)\rangle  = \sum_i\langle n_i(t)n_i(0)\rangle + \sum_{i, j\neq i}\langle n_j(t)n_i(0)\rangle \,\,.
\label{eq:allnumcorr}
\end{equation}

For non-interacting systems particles are independent and Eq.~\eqref{eq:allnumcorr} can be rewritten as \cite{SuppMat} 
\begin{equation}
\label{eq:Pin}
   \langle N(t)N(0)\rangle=\meanN P_{\rm in}(t) +\meanN^2,  
\end{equation}
where $P_{\rm in}(t)$ is the probability that a particle which starts in the box at $t=0$ is ``in'' the box  at time $t$, either because it remained in or returned to the box. 
The high frequency $\sim f^{-3/2}$ regime corresponds to the early time behaviour of $P_{\rm in}(t)$. Over short timescales $t$, particles escape the box in bands of width $l\sim\sqrt{Dt}$ from the box edges, resulting in a $\sqrt{t}$ dependence for $P_{\rm in}(t)$ \cite{Marbach2021,Mackay2024} which translates by Fourier transform to $S(f)\sim f^{-3/2}$. 

Over longer time scales, all particles initially in the box will have diffused out. As such, correlation in $P_{\rm in}(t)$ can only arise if a particle returns to the box it initially occupied. For a 3D system, it has long been recognised that the probability of such a return vanishes at long timescales. For diffusion in 2D as considered here, however, $\Pin$ decays over long times as $1/t$, precisely as $L^2/4\pi Dt$, and so there is always a finite probability that a particle returns to its initial box. These time-dependent particle returns are probed in both small and large boxes, hence the collapse of data for multiple box sizes to the same low-frequency $\sim\log(f)$ in Fig.~\ref{fig:fig1}(b); the box size only affects the time scale at which this regime begins.

\begin{figure}
    \centering
    \includegraphics[width=0.95\linewidth]{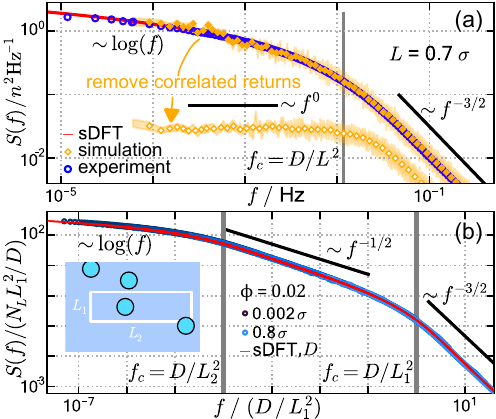}
    \caption{\textbf{Persistent particle returns in dilute systems induce low-frequency noise.}
    (a) N-PSD for a single box with $L=0.7\sigma$ from experiment (blue) and Brownian Dynamics simulations (yellow) with (open) and without (closed)  artificial particle repositioning.
    (b) N-PSDs for particle counts in rectangular boxes of fixed aspect ratio ($L_2/L_1=100)$ and variable size from $L_1=0.002\,\sigma$ (dark blue) to $L_1=0.8\,\sigma$ (light blue). Points show experimental data; red line shows sDFT theory (without interactions) \cite{SuppMat}. }
    \label{fig:return}
\end{figure}

To explicitly demonstrate the effect of particle returns on $S(f)$, in Fig.~\ref{fig:return}(a) we show a comparison between experimental data and results from Brownian Dynamics simulations. Here, the true 2D simulation (closed yellow symbols) perfectly reproduces the experimental spectra. If a particle is randomly repositioned within the simulation domain upon leaving the box, however, there is an immediate loss of the low-frequency $-\log(f)$ scaling (open yellow symbols) \cite{DrummondYoung2025}. 
The fact that low-frequency scalings link to return probability also introduces a striking dependence on dimensionality. This is illustrated in Fig.~\ref{fig:return}(b) which shows spectra for a series of rectangular boxes of fixed aspect ratio ($L_2/L_1=100$). Here, at intermediate frequencies, we probe flux predominantly through the long edges, \textit{i.e.} crossing only the shortest axis of the box, and resolve a $f^{-1/2}$ scaling characteristic of 1D systems \cite{Burgess1965,Voss1976,Marbach2021}. This transitions to the expected 2D behaviour at lower frequencies, at times long enough to sense flux over both dimensions. These scaling laws are consistent with analytical results from solid-state physics \cite{lax1960influence,Voss1976,Burgess1965,Bezrukov2000,berezhkovskii2002effect} as well as more recent theory and simulations of soft systems \cite{Marbach2021,Minh2023,DrummondYoung2025}.  
Notably, however, we now present a realisation of both the high- and low-frequency noise regimes in an experimental platform. 

\begin{figure}
    \centering
    \includegraphics[width=0.9\linewidth]{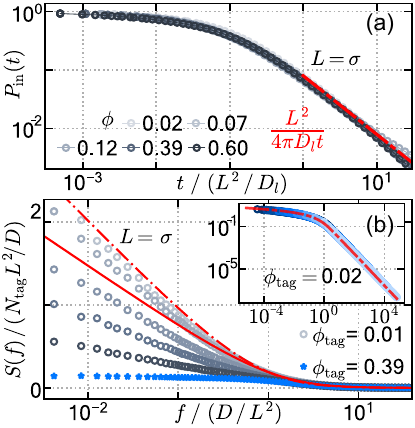}
    \caption{\textbf{ Single particle return probabilities do not qualitatively change with density.} (a) Experimental $P_{\mathrm{in}}(t)$ at different $\phi$ for a small box at $L=\sigma$ with the predicted limiting slope $\Pin=L^2/4\pi D_l t$ (red dot-dashed line).
    (b) PSD of $N_{\rm tag}(t)$ for particles randomly tagged at density $\phi_{\rm tag}$ in the system at total $\phi = 0.39$. Points show experimental data with increasing percentage of tagged particles from $2~\%$ (pale grey) to $100~\%$ (blue stars); Red lines show sDFT theory (without interactions) using $\Dshort$ (solid line) or $D_l$ (dot-dashed).
    Inset shows the corresponding experimental data for fixed $\phi_{\rm tag}=0.02$ and varying $L=\{0.1,0.2,0.5,1.0,1.9,3.8,7.6,15.3\}\,\sigma$, rescaled as in Fig.~\ref{fig:fig1}(b). (Red line) sDFT theory without interactions using $D_l$ \cite{SuppMat}.
    }
    \label{fig:fig3}
\end{figure}

Intuitively, interactions would modify particle return probabilities. As such, to explore the origins of the spectral changes in Fig.~\ref{fig:fig1}(c) we first consider how $\Pin$ varies with $\phi$ for a small fixed box size.  
We directly calculate $P_{\rm in}(t)$ from experimental particle trajectories, as shown in Fig.~\ref{fig:fig3}(a). We find perfect collapse to one curve at long times if time is rescaled by $L^2/D_l$ with the \textit{long time} self-diffusion coefficient, $D_l = \lim_{t\rightarrow\infty }\langle r^2(t)\rangle/4t$. Additionally, all data tend asymptotically to the expected limit $P_{\rm in}^{\infty}(t) =L^2/4\pi D_l t$.  This shows that qualitative changes in the N-PSD arising from interactions are not a consequence of changes in individual particle returns.  

To further confirm this, we perform alternative measurements where only a few randomly chosen particles are ``tagged'' and contribute to the counts, $N_{\rm tag}(t)$.  When the fraction of tagged particles $\phi_{\rm tag}$ is low enough, we can recover ``non-interacting"-like N-PSDs from the dense experimental data, which agree perfectly with the non-interacting sDFT theory using $D_l$ as input. This is shown in Fig.~\ref{fig:fig3}(b, inset), where we plot rescaled spectra for multiple box sizes for $\phi=0.39$, but with tagged packing fraction $\phi_{\rm tag} = 0.02$. Notably, the low frequency $-\log(f)$ scaling implies that particles return to their point of origin with the same time-dependence, even though they are now exploring a dense environment.
Tagged particle spectra are, however, very sensitive to the proportion of tagged particles. Fig.~\ref{fig:fig3}(b) shows a steady transition in the low-frequency scaling  as $\phi_{\rm tag}$ is increased for a small fixed box size, with the dilute $S(f)\sim -{\rm log}(f)$ scaling fully suppressed for $\phi_{\rm tag}=\phi$. We note that this tagged particle measurement resembles the approach of FCS, where only a fraction of fluorescently-labelled particles contribute to the intensity signal \cite{Hofling2011}. Our results highlight the critical importance of using a sufficiently small tagged particle fraction to accurately apply non-interacting fits to interacting systems. 

As the spectral signatures of number fluctuations in dense systems cannot be rationalised by single particle behaviour alone, 
we now reconsider the second term of Eq.~\eqref{eq:allnumcorr}, which captures correlations between different particles. 
Converting to a description via the probability, $P$, to observe specific configurations \cite{SuppMat}, 
\begin{equation*}
\begin{split}
    \sum_{i,j\neq i}&\langle n_i(0)n_j(t)\rangle  \\
     &= \sum_iP(n_i(0)=1)\sum_{j\neq i}P(n_j(t)=1\,|\, n_i(0)=1) \\
    &=  \meanN \sum_{j\neq b}P(n_j(t)=1\,|\, n_b(0)=1) \equiv \meanN \Nnew\,\, ,
\end{split}
\end{equation*}
where the final sum of probabilities, $\Nnew$, represents the mean number of ``other" particles that are in the box at $t$ given some tracer particle, $b$, was in the box at $t=0$. 
Eq.~\eqref{eq:Pin} for interacting systems now becomes:
\begin{equation}
\label{eq:finalcorrfunc}
    \langle N(t)N(0)\rangle = \meanN P_{\rm in}(t) + \meanN\Nnew\,\,  
\end{equation} 
suggesting that qualitative changes in spectra at high $\phi$ arise from $\Nnew$. Note that for low $\phi$, $\Nnew \simeq \meanN$ recovering Eq.~\eqref{eq:Pin}. Unlike $\Pin$, this term probes different transport processes depending on the box size. In small boxes ($L\lesssim\sigma$), only one incoming particle can contribute to $\Nnew$ while in large boxes ($L\gg\sigma$) many particles can contribute. We therefore probe pairwise replacement in small boxes versus collective rearrangements in large boxes.

\begin{figure}[h!]
    \centering
    \includegraphics[width=0.95\linewidth]{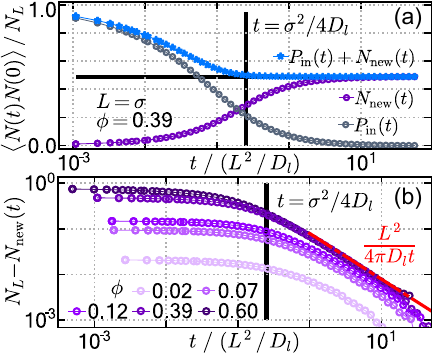}
    \caption{\textbf{Hard core interactions mediate anti-correlations between pairs of particles.}
    (a) Single particle ($\Pin$, grey points), interparticle ($\Nnew$, purple points) and total ($\Pin + \Nnew$, blue stars) correlations in a dense ($\phi=0.39$) experimental system for $L=\,\sigma$. (b) For all experimental packing fractions, $\Nnew$ tends to $\meanN$ at long time as $1/t$.
    }
    \label{fig:fig4}
\end{figure}

In Fig.~\ref{fig:fig4}(a) we show a direct calculation of $\Pin$, $\Nnew$ and their sum for a small box ($L=\sigma$) at $\phi=0.39$. Data show that $\Nnew$ indeed has non-trivial time dependence, growing from zero at short times to approach the mean particle number, $\meanN$,  for long times. Crucially, the growth of $\Nnew$ over time is found to have an identical, but opposite, scaling to the decay in $\Pin$ at long times. 
This can be seen in Fig.~\ref{fig:fig4}(b) where we show that $\meanN - \Nnew$ for different packing fractions systematically reaches the limiting scaling of $L^2/4\pi D_l t$. 
This coupling between $\Pin$ and $\Nnew$ in small boxes creates a regime where $\langle N(t)N(0)\rangle-\meanN^2$ decays more rapidly than $1/t$, which creates a plateau in the N-PSD. The complete set of scalings displayed in the temporal correlation functions are shown in the SI \cite{SuppMat}.

We rationalise this mechanistically as follows. At short times, a particle fills any sufficiently small box and thus $\Nnew=0$. Over time, diffusion of the initial particle away from the box allows new particles to enter and $\Nnew$ grows. Clearly, a new particle can only move into the space when the original particle has diffused a distance on the order of the particle diameter. This makes the relevant timescale $t\simeq\sigma^2/\Dshort \simeq \sigma^2/D_l$ (Fig.~\ref{fig:fig1}(c)). For long times, $\Nnew \rightarrow \meanN$ as memory of the initial box occupant is progressively lost. The rate that this is lost, however, corresponds to return kinetics of any \textit{other} particle to the box, giving us a $1/t$ scaling in~$\Nnew$.

At low $\phi$, while the same trends persist (see SI \cite{SuppMat}), we only observe scalings related to $\Pin$ in the spectra. In these low density cases the decay $\meanN-\Nnew$ is negligible compared to $\Pin$. Physically, this is because at low $\phi$, few boxes are ``blocked'' by particles and so the total correlation function is dominated by self-particle returns. 

While pairwise particle replacements persist in large boxes, we do not see spectral plateaus (see SI \cite{SuppMat} for a direct comparison of the N-PSD in large boxes at multiple packing fractions). This is because fluctuations in $N(t)$ are dominated by collective rearrangements involving multiple particles, \textit{i.e}, the relaxation of $\Nnew$ is governed by collective diffusion. 
The $-\log(f)$ regime in large boxes at high $\phi$ can thus be used to experimentally probe the collective diffusion $D_{\rm coll}$. Physically, the $S(f)\sim -\log(f)$ scaling in this frequency regime means we have accessed the time and length scales over which the many-body system is effectively a continuum fluid, described with a number density $\rho$, where fluctuations in this number density relax diffusively \cite{Dhont1996,Carter2025}. Interestingly, analysis of time correlation functions \cite{SuppMat} suggests this scaling regime would also arise in small box N-PSDs but at frequencies lower than those accessible to experiment.

Our results demonstrate that the power spectrum of number fluctuations naturally probes a hierarchy of self, pairwise, and collective particle dynamics, simply by considering different length and time scales within this one correlation function. The N-PSD applies to dense interacting systems with comparative ease,  and reveals details of dynamics that are difficult to extract directly from particle trajectories. Moreover, without a priori knowledge of the structure factor, we can extract the range of the interaction potential, here $\sigma$, from the corner frequency in small-box spectra. Despite the apparent simplicity of this foundational interacting system, the N-PSD exposes subtle transport mechanisms in hard spheres, including long time return kinetics, a feature of low dimensional systems, and non-trivial collective behaviour. Notably, a plateau in the spectrum -- though commonly interpreted as the absence of correlations -- emerges through the interplay of two correlated processes. 

Our straightforward analysis offers varied opportunities in more complex systems -- including longer-ranged repulsive and attractive potentials, the effect of which can be included in the analytical theory if the structure factor $S(k)$ is known. For example, by analogy with variable length scale Fluorescence Correlation Spectroscopy, the N-PSD cound be extended to distinguish distinct subdiffusive or anomalous random walks~\cite{Hofling2011,Banks2016}. Sensitivity to self and collective dynamics should make the N-PSD useful for probing active matter, where self-diffusion can be enhanced \cite{Howse2007} and collective transport includes additional modes of motion \cite{Vicsek1995,Wang2011spontaneous,Mognetti2013,Narayan2007}. Further insights could be gained by extending to higher-order correlation functions, \textit{e.g.} number analogues of the four-point correlation functions used to probe dynamical heterogeneity in glasses \cite{pastore2022multiscale,wang2012brownian,Berthier2011}. Finally, varying box size, shape and location enables dynamics to be probed in systems with complex boundaries, such as in porous media. Spectral analysis is already commonly used to probe nanopore transport phenomena, although a lack of single-particle information combined with the complexity of molecular mechanisms often make physical interpretation challenging \cite{smeets2008noise,smeets2009low,siwy2002origin,Zorkot2016}. Given that reservoir fluctuations are now thought to play a governing role in nanopore transport \cite{Gravelle2019,fragasso20191,Robin2023}, this could be explored further by employing moveable box counting in simulations or microscale model experiments \cite{knowles2024interpreting}.



\begin{acknowledgements}
The authors are grateful for fruitful discussions with Kurt Andresen, Tristan Cerdin, Roel Dullens and Brennan Sprinkle. 
E. K. R. M. and A. L. T. acknowledge funding from EPSRC (EP/X02492X/1).  A. D. Y. acknowledges funding from EPSRC (EP/W524311/1). A. L. T. acknowledges funding from a Royal Society University Research Fellowship (URF/R1/211033). 
A.C. acknowledges funding from Institute of Materials Science of the Alliance Sorbonne Université iMAT for a PhD grant. 
\end{acknowledgements}


\bibliography{bibliography}

@article{metzler2019brownian,
  title={Brownian motion and beyond: first-passage, power spectrum, non-Gaussianity, and anomalous diffusion},
  author={Metzler, Ralf},
  journal={Journal of Statistical Mechanics: Theory and Experiment},
  volume={2019},
  number={11},
  pages={114003},
  year={2019},
  publisher={IOP Publishing}
}

@book{dhont1996introduction,
  title={An introduction to dynamics of colloids},
  author={Dhont, Jan KG},
  volume={2},
  year={1996},
  publisher={Elsevier}
}

@article{lattuada2025hitchhiker,
  title={The hitchhiker’s guide to differential dynamic microscopy},
  author={Lattuada, Enrico and Krautgasser, Fabian and Lavaud, Maxime and Giavazzi, Fabio and Cerbino, Roberto},
  journal={The Journal of Chemical Physics},
  volume={163},
  number={16},
  year={2025},
  publisher={AIP Publishing}
}

@article{moisan2011periodic,
  title={Periodic plus smooth image decomposition},
  author={Moisan, Lionel},
  journal={Journal of Mathematical Imaging and Vision},
  volume={39},
  pages={161--179},
  year={2011},
  publisher={Springer}
}

@article{giavazzi2017image,
  title={Image windowing mitigates edge effects in Differential Dynamic Microscopy},
  author={Giavazzi, Fabio and Edera, Paolo and Lu, Peter J and Cerbino, Roberto},
  journal={The European Physical Journal E},
  volume={40},
  pages={1--9},
  year={2017},
  publisher={Springer}
}

@article{zuccolotto2024improving,
  title={Improving data sampling with rapid statistical convergence in digital Fourier microscopy analysis},
  author={Zuccolotto-Bernez, AB and Rojas-Ochoa, LF and Egelhaaf, SU and Escobedo-S{\'a}nchez, MA},
  journal={Applied Optics},
  volume={63},
  number={34},
  pages={8760--8770},
  year={2024},
  publisher={Optica Publishing Group},
url = {https://opg.optica.org/ao/abstract.cfm?URI=ao-63-34-8760},
doi = {10.1364/AO.537840},
}

@article{manzo2015review,
  title={A review of progress in single particle tracking: from methods to biophysical insights},
  author={Manzo, Carlo and Garcia-Parajo, Maria F},
  journal={Reports on progress in physics},
  volume={78},
  number={12},
  pages={124601},
  year={2015},
  publisher={IOP Publishing}
}

@article{rose2020particle,
  title={Particle tracking of nanoparticles in soft matter},
  author={Rose, Katie A and Molaei, Mehdi and Boyle, Michael J and Lee, Daeyeon and Crocker, John C and Composto, Russell J},
  journal={Journal of Applied Physics},
  volume={127},
  number={19},
  year={2020},
  publisher={AIP Publishing}
}

@article{Henderson2002,
  title = {Propagation of Hydrodynamic Interactions in Colloidal Suspensions},
  author = {Henderson, Stuart and Mitchell, Steven and Bartlett, Paul},
  journal = {Phys. Rev. Lett.},
  volume = {88},
  issue = {8},
  pages = {088302},
  numpages = {4},
  year = {2002},
  month = {Feb},
  publisher = {American Physical Society},
  doi = {10.1103/PhysRevLett.88.088302},
  url = {https://link.aps.org/doi/10.1103/PhysRevLett.88.088302}
}

@article{Franosch2011,
  title={Resonances arising from hydrodynamic memory in Brownian motion},
  author={Franosch, Thomas and Grimm, Matthias and Belushkin, Maxim and Mor, Flavio M and Foffi, Giuseppe and Forr{\'o}, L{\'a}szl{\'o} and Jeney, Sylvia},
  journal={Nature},
  volume={478},
  number={7367},
  pages={85--88},
  year={2011},
  publisher={Nature Publishing Group UK London}
}

@article{Bezrukov2000,
    author = {Bezrukov, Sergey M. and Berezhkovskii, Alexander M. and Pustovoit, Mark A. and Szabo, Attila},
    title = {Particle number fluctuations in a membrane channel},
    journal = {The Journal of Chemical Physics},
    volume = {113},
    number = {18},
    pages = {8206-8211},
    year = {2000},
    month = {11},
    issn = {0021-9606},
    doi = {10.1063/1.1314862},
    url = {https://doi.org/10.1063/1.1314862},
    eprint = {https://pubs.aip.org/aip/jcp/article-pdf/113/18/8206/19288885/8206\_1\_online.pdf},
}

@inbook{Burgess1965,
  title={Fluctuation phenomena in solids},
  author={K. M. van Vliet and J. R. Fassett},
  editor={Burgess, Ronald Eric},
  year={1965},
  publisher={American Institute of Physics},
  chapter={Fluctuations due to electronic transitions and transport in solids},
  pages={267-354}
}

@article{Voss1976,
  title = {Flicker ($\frac{1}{f}$) noise: Equilibrium temperature and resistance fluctuations},
  author = {Voss, Richard F. and Clarke, John},
  journal = {Phys. Rev. B},
  volume = {13},
  issue = {2},
  pages = {556--573},
  numpages = {0},
  year = {1976},
  month = {Jan},
  publisher = {American Physical Society},
  doi = {10.1103/PhysRevB.13.556},
  url = {https://link.aps.org/doi/10.1103/PhysRevB.13.556}
}

@article{Robin2023,
author ="Robin, Paul and Lizée, Mathieu and Yang, Qian and Emmerich, Théo and Siria, Alessandro and Bocquet, Lydéric",
title  ="Disentangling 1/f noise from confined ion dynamics",
journal  ="Faraday Discuss.",
year  ="2023",
volume  ="246",
issue  ="0",
pages  ="556-575",
publisher  ="The Royal Society of Chemistry",
doi  ="10.1039/D3FD00035D",
url  ="http://dx.doi.org/10.1039/D3FD00035D"
}

@article{Minton2006,
  title={How can biochemical reactions within cells differ from those in test tubes?},
  author={Minton, Allen P},
  journal={Journal of cell science},
  volume={119},
  number={14},
  pages={2863--2869},
  year={2006},
  publisher={Company of Biologists}
}

@article{Szymanski2009,
  title={Elucidating the origin of anomalous diffusion in crowded fluids},
  author={Szymanski, Jedrzej and Weiss, Matthias},
  journal={Physical review letters},
  volume={103},
  number={3},
  pages={038102},
  year={2009},
  publisher={APS}
}

@article{Waigh2023,
  title={Heterogeneous anomalous transport in cellular and molecular biology},
  author={Waigh, Thomas Andrew and Korabel, Nickolay},
  journal={Reports on Progress in Physics},
  volume={86},
  number={12},
  pages={126601},
  year={2023},
  publisher={IOP Publishing}
}

@article{Weiss2004,
  title={Anomalous subdiffusion is a measure for cytoplasmic crowding in living cells},
  author={Weiss, Matthias and Elsner, Markus and Kartberg, Fredrik and Nilsson, Tommy},
  journal={Biophysical journal},
  volume={87},
  number={5},
  pages={3518--3524},
  year={2004},
  publisher={Elsevier}
}

@article{Golding2006,
  title={Physical nature of bacterial cytoplasm},
  author={Golding, Ido and Cox, Edward C},
  journal={Physical review letters},
  volume={96},
  number={9},
  pages={098102},
  year={2006},
  publisher={APS}
}

@article{Hofling2013,
  title={Anomalous transport in the crowded world of biological cells},
  author={H{\"o}fling, Felix and Franosch, Thomas},
  journal={Reports on Progress in Physics},
  volume={76},
  number={4},
  pages={046602},
  year={2013},
  publisher={IOP Publishing}
}

@Article{Banks2016,
author ="Banks, Daniel S. and Tressler, Charmaine and Peters, Robert D. and Höfling, Felix and Fradin, Cécile",
title  ="Characterizing anomalous diffusion in crowded polymer solutions and gels over five decades in time with variable-lengthscale fluorescence correlation spectroscopy",
journal  ="Soft Matter",
year  ="2016",
volume  ="12",
issue  ="18",
pages  ="4190-4203",
publisher  ="The Royal Society of Chemistry",
doi  ="10.1039/C5SM01213A"
}

@article{Saxton2007,
  title={A biological interpretation of transient anomalous subdiffusion. I. Qualitative model},
  author={Saxton, Michael J},
  journal={Biophysical journal},
  volume={92},
  number={4},
  pages={1178--1191},
  year={2007},
  publisher={Elsevier}
}

@article{Digman2011,
  title={Lessons in fluctuation correlation spectroscopy},
  author={Digman, Michelle A and Gratton, Enrico},
  journal={Annual review of physical chemistry},
  volume={62},
  number={1},
  pages={645--668},
  year={2011},
  publisher={Annual Reviews}
}

@article{Tolic2004,
  title={Anomalous diffusion in living yeast cells},
  author={Toli{\'c}-N{\o}rrelykke, Iva Marija and Munteanu, Emilia-Laura and Thon, Genevieve and Oddershede, <? format?> Lene and Berg-S{\o}rensen, Kirstine},
  journal={Physical review letters},
  volume={93},
  number={7},
  pages={078102},
  year={2004},
  publisher={APS}
}

@article{Krapf2018,
doi = {10.1088/1367-2630/aaa67c},
url = {https://dx.doi.org/10.1088/1367-2630/aaa67c},
year = {2018},
month = {feb},
publisher = {IOP Publishing},
volume = {20},
number = {2},
pages = {023029},
author = {Diego Krapf and Enzo Marinari and Ralf Metzler and Gleb Oshanin and Xinran Xu and Alessio Squarcini},
title = {Power spectral density of a single Brownian trajectory: what one can and cannot learn from it},
journal = {New Journal of Physics}
}

@article{Krapf2019,
  title={Spectral content of a single non-Brownian trajectory},
  author={Krapf, Diego and Lukat, Nils and Marinari, Enzo and Metzler, Ralf and Oshanin, Gleb and Selhuber-Unkel, Christine and Squarcini, Alessio and Stadler, Lorenz and Weiss, Matthias and Xu, Xinran},
  journal={Physical Review X},
  volume={9},
  number={1},
  pages={011019},
  year={2019},
  publisher={APS}
}

@article{Sposini2019,
  title={Single-trajectory spectral analysis of scaled Brownian motion},
  author={Sposini, Vittoria and Metzler, Ralf and Oshanin, Gleb},
  journal={New Journal of Physics},
  volume={21},
  number={7},
  pages={073043},
  year={2019},
  publisher={IOP Publishing}
}

@article{Fox2021,
  title={Aging power spectrum of membrane protein transport and other subordinated random walks},
  author={Fox, Zachary R and Barkai, Eli and Krapf, Diego},
  journal={Nature communications},
  volume={12},
  number={1},
  pages={6162},
  year={2021},
  publisher={Nature Publishing Group UK London}
}

@article{Sposini2022,
  title={Towards a robust criterion of anomalous diffusion},
  author={Sposini, Vittoria and Krapf, Diego and Marinari, Enzo and Sunyer, Raimon and Ritort, Felix and Taheri, Fereydoon and Selhuber-Unkel, Christine and Benelli, Rebecca and Weiss, Matthias and Metzler, Ralf and others},
  journal={Communications Physics},
  volume={5},
  number={1},
  pages={305},
  year={2022},
  publisher={Nature Publishing Group UK London}
}

@article{Thorneywork2014,
    author = {Thorneywork, Alice L. and Roth, Roland and Aarts, Dirk G. A. L. and Dullens, Roel P. A.},
    title = {Communication: Radial distribution functions in a two-dimensional binary colloidal hard sphere system},
    journal = {The Journal of Chemical Physics},
    volume = {140},
    number = {16},
    pages = {161106},
    year = {2014},
    month = {04},
    doi = {10.1063/1.4872365},
    url = {https://doi.org/10.1063/1.4872365},
    eprint = {https://pubs.aip.org/aip/jcp/article-pdf/doi/10.1063/1.4872365/14007296/161106\_1\_online.pdf},
}

@article{Crocker1996,
  title={Methods of digital video microscopy for colloidal studies},
  author={Crocker, John C and Grier, David G},
  journal={J. Colloid Interface Sci.},
  volume={179},
  number={1},
  pages={298},
  year={1996},
  publisher={Elsevier},
  doi = {https://doi.org/10.1006/jcis.1996.0217}
}

@software{Trackpy2023,
  author       = {Allan, Daniel B. and
                  Caswell, Thomas and
                  Keim, Nathan C. and
                  van der Wel, Casper M. and
                  Verweij, Ruben W.},
  title        = {soft-matter/trackpy: v0.6.1},
  month        = feb,
  year         = 2023,
  publisher    = {Zenodo},
  version      = {v0.6.1},
}

@book{Press1992,
    author    = {Press, William H. and
		Teukolsky, Saul A. and
		Vetterling, William T. and
		Flannery, Brian P.},
    title     = {Numerical Recipes in C: The Art of Scientific Computing},
    publisher = {Cambridge University Press},
    year      = 1992
}

@book{Bendat2011,
  title={Random data: analysis and measurement procedures},
  author={Bendat, Julius S and Piersol, Allan G},
  year={2011},
  publisher={John Wiley \& Sons}
}

@article{Marbach2021,
    author = {Marbach, S.},
    title = {Intrinsic fractional noise in nanopores: The effect of reservoirs},
    journal = {The Journal of Chemical Physics},
    volume = {154},
    number = {17},
    pages = {171101},
    year = {2021},
    month = {05},
    issn = {0021-9606},
    doi = {10.1063/5.0047380},
    url = {https://doi.org/10.1063/5.0047380},
    eprint = {https://pubs.aip.org/aip/jcp/article-pdf/doi/10.1063/5.0047380/13534122/171101\_1\_online.pdf},
}

@article{Mackay2024,
  title = {The Countoscope: Measuring Self and Collective Dynamics without Trajectories},
  author = {Mackay, Eleanor K. R. and Marbach, Sophie and Sprinkle, Brennan and Thorneywork, Alice L.},
  journal = {Phys. Rev. X},
  volume = {14},
  issue = {4},
  pages = {041016},
  numpages = {14},
  year = {2024},
  month = {Oct},
  publisher = {American Physical Society},
  doi = {10.1103/PhysRevX.14.041016},
  url = {https://link.aps.org/doi/10.1103/PhysRevX.14.041016}
}

@article{DrummondYoung2025,
    author = {Drummond Young, Anna and Thorneywork, A. L. and Marbach, S.},
    title = {Decoding noise in nanofluidic systems: Adsorption vs diffusion signatures in power spectra},
    journal = {The Journal of Chemical Physics},
    volume = {163},
    number = {21},
    pages = {214902},
    year = {2025},
    month = {12},
    issn = {0021-9606},
    doi = {10.1063/5.0288288},
}

@Article{Minh2023,
author ="Hoang Ngoc Minh, Thê and Rotenberg, Benjamin and Marbach, Sophie",
title  ="Ionic fluctuations in finite volumes: fractional noise and hyperuniformity",
journal  ="Faraday Discuss.",
year  ="2023",
volume  ="246",
issue  ="0",
pages  ="225-250",
publisher  ="The Royal Society of Chemistry",
doi  ="10.1039/D3FD00031A",
url  ="http://dx.doi.org/10.1039/D3FD00031A"
}

@book{Dhont1996,
  title={An introduction to dynamics of colloids},
  author={Dhont, Jan KG},
  volume={2},
  year={1996},
  publisher={Elsevier}
}

@article{Zorkot2016,
  title={The power spectrum of ionic nanopore currents: the role of ion correlations},
  author={Zorkot, Mira and Golestanian, Ramin and Bonthuis, Douwe Jan},
  journal={Nano letters},
  volume={16},
  number={4},
  pages={2205--2212},
  year={2016},
  publisher={ACS Publications}
}

@article{pastore2022multiscale,
  title={Multiscale heterogeneous dynamics in two-dimensional glassy colloids},
  author={Pastore, Raffaele and Giavazzi, Fabio and Greco, Francesco and Cerbino, Roberto},
  journal={The Journal of Chemical Physics},
  volume={156},
  number={16},
  year={2022},
  publisher={AIP Publishing}
}

@article{lu2012characterizing,
   title = {Characterizing Concentrated, Multiply Scattering, and Actively Driven Fluorescent Systems with Confocal Differential Dynamic Microscopy},
  author = {Lu, Peter J. and Giavazzi, Fabio and Angelini, Thomas E. and Zaccarelli, Emanuela and Jargstorff, Frank and Schofield, Andrew B. and Wilking, James N. and Romanowsky, Mark B. and Weitz, David A. and Cerbino, Roberto},
  journal = {Phys. Rev. Lett.},
  volume = {108},
  issue = {21},
  pages = {218103},
  numpages = {5},
  year = {2012},
  month = {May},
  publisher = {American Physical Society},
  doi = {10.1103/PhysRevLett.108.218103},
  url = {https://link.aps.org/doi/10.1103/PhysRevLett.108.218103}
}

@article{wang2012brownian,
  title={When Brownian diffusion is not Gaussian},
  author={Wang, Bo and Kuo, James and Bae, Sung Chul and Granick, Steve},
  journal={Nature materials},
  volume={11},
  number={6},
  pages={481--485},
  year={2012},
  publisher={Nature Publishing Group UK London}
}

@article{berezhkovskii2002effect,
  title={Effect of binding on particle number fluctuations in a membrane channel},
  author={Berezhkovskii, Alexander M and Pustovoit, Mark A and Bezrukov, Sergey M},
  journal={The Journal of chemical physics},
  volume={116},
  number={14},
  pages={6216--6220},
  year={2002},
  publisher={American Institute of Physics}
}

@article{knowles2024interpreting,
  title={Interpreting the power spectral density of a fluctuating colloidal current},
  author={Knowles, Stuart F and Mackay, Eleanor KR and Thorneywork, Alice L},
  journal={The Journal of Chemical Physics},
  volume={161},
  number={14},
  year={2024},
  publisher={AIP Publishing}
}

@article{smeets2008noise,
  title={Noise in solid-state nanopores},
  author={Smeets, Ralph MM and Keyser, Ulrich F and Dekker, Nynke H and Dekker, Cees},
  journal={Proceedings of the National Academy of Sciences},
  volume={105},
  number={2},
  pages={417--421},
  year={2008},
  publisher={National Academy of Sciences}
}

@article{smeets2009low,
  title={Low-frequency noise in solid-state nanopores},
  author={Smeets, RMM and Dekker, NH and Dekker, C},
  journal={Nanotechnology},
  volume={20},
  number={9},
  pages={095501},
  year={2009},
  publisher={IOP Publishing}
}

@article{Carter2025,
  title={Measuring collective diffusion coefficients by counting particles in boxes},
  author={Carter, Adam and Mackay, Eleanor KR and Sprinkle, Brennan and Thorneywork, Alice L and Marbach, Sophie},
  journal={Soft Matter},
  volume={21},
  number={20},
  pages={3991--4002},
  year={2025},
  publisher={Royal Society of Chemistry}
}

@article{fragasso20191,
  title={1/f noise in solid-state nanopores is governed by access and surface regions},
  author={Fragasso, Alessio and Pud, Sergii and Dekker, Cees},
  journal={Nanotechnology},
  volume={30},
  number={39},
  pages={395202},
  year={2019},
  publisher={IOP Publishing}
}

@article{Gravelle2019,
  title={Adsorption kinetics in open nanopores as a source of low-frequency noise},
  author={Gravelle, Simon and Netz, Roland R and Bocquet, Lyderic},
  journal={Nano letters},
  volume={19},
  number={10},
  pages={7265--7272},
  year={2019},
  publisher={ACS Publications}
}

@article{dean1996langevin,
  title={Langevin equation for the density of a system of interacting Langevin processes},
  author={Dean, David S},
  journal={Journal of Physics A: Mathematical and General},
  volume={29},
  number={24},
  pages={L613},
  year={1996},
  publisher={IOP Publishing}
}

@article{lax1960influence,
  title={Influence of trapping, diffusion and recombination on carrier concentration fluctuations},
  author={Lax, M and Mengert, P},
  journal={Journal of Physics and Chemistry of Solids},
  volume={14},
  pages={248--267},
  year={1960},
  publisher={Elsevier}
}

@article{Alexandre2023,
  title = {Non-Gaussian Diffusion Near Surfaces},
  author = {Alexandre, Arthur and Lavaud, Maxime and Fares, Nicolas and Millan, Elodie and Louyer, Yann and Salez, Thomas and Amarouchene, Yacine and Gu\'erin, Thomas and Dean, David S.},
  journal = {Phys. Rev. Lett.},
  volume = {130},
  issue = {7},
  pages = {077101},
  numpages = {7},
  year = {2023},
  month = {Feb},
  publisher = {American Physical Society},
  doi = {10.1103/PhysRevLett.130.077101},
}

@article{Guan2014,
author = {Guan, Juan and Wang, Bo and Granick, Steve},
title = {Even Hard-Sphere Colloidal Suspensions Display Fickian Yet Non-Gaussian Diffusion},
journal = {ACS Nano},
volume = {8},
number = {4},
pages = {3331-3336},
year = {2014},
doi = {10.1021/nn405476t}
}

@article{Wang2009,
author = {Bo Wang  and Stephen M. Anthony  and Sung Chul Bae  and Steve Granick },
title = {Anomalous yet Brownian},
journal = {Proceedings of the National Academy of Sciences},
volume = {106},
number = {36},
pages = {15160-15164},
year = {2009},
doi = {10.1073/pnas.0903554106}
}

@article{Skinner2013,
  title = {Localization Dynamics of Fluids in Random Confinement},
  author = {Skinner, Thomas O. E. and Schnyder, Simon K. and Aarts, Dirk G. A. L. and Horbach, J\"urgen and Dullens, Roel P. A.},
  journal = {Phys. Rev. Lett.},
  volume = {111},
  issue = {12},
  pages = {128301},
  numpages = {5},
  year = {2013},
  month = {Sep},
  publisher = {American Physical Society},
  doi = {10.1103/PhysRevLett.111.128301}
}

@article{Sentjabrskaja2016,
  title={Anomalous dynamics of intruders in a crowded environment of mobile obstacles},
  author={Sentjabrskaja, Tatjana and Zaccarelli, Emanuela and De Michele, Cristiano and Sciortino, Francesco and Tartaglia, Piero and Voigtmann, Thomas and Egelhaaf, Stefan U and Laurati, Marco},
  journal={Nature communications},
  volume={7},
  number={1},
  pages={11133},
  year={2016},
  publisher={Nature Publishing Group UK London}
}

@article{Pastore2021,
  title = {Rapid Fickian Yet Non-Gaussian Diffusion after Subdiffusion},
  author = {Pastore, Raffaele and Ciarlo, Antonio and Pesce, Giuseppe and Greco, Francesco and Sasso, Antonio},
  journal = {Phys. Rev. Lett.},
  volume = {126},
  issue = {15},
  pages = {158003},
  numpages = {6},
  year = {2021},
  month = {Apr},
  publisher = {American Physical Society},
  doi = {10.1103/PhysRevLett.126.158003}
}

@article{Rusciano2022,
  title = {Fickian Non-Gaussian Diffusion in Glass-Forming Liquids},
  author = {Rusciano, Francesco and Pastore, Raffaele and Greco, Francesco},
  journal = {Phys. Rev. Lett.},
  volume = {128},
  issue = {16},
  pages = {168001},
  numpages = {6},
  year = {2022},
  month = {Apr},
  publisher = {American Physical Society},
  doi = {10.1103/PhysRevLett.128.168001}
}

@article{Hofling2011,
author ="Höfling, Felix and Bamberg, Karl-Ulrich and Franosch, Thomas",
title  ="Anomalous transport resolved in space and time by fluorescence correlation spectroscopy",
journal  ="Soft Matter",
year  ="2011",
volume  ="7",
issue  ="4",
pages  ="1358-1363",
publisher  ="The Royal Society of Chemistry",
doi  ="10.1039/C0SM00718H"
}

@article{Howse2007,
  title = {Self-Motile Colloidal Particles: From Directed Propulsion to Random Walk},
  author = {Howse, Jonathan R. and Jones, Richard A. L. and Ryan, Anthony J. and Gough, Tim and Vafabakhsh, Reza and Golestanian, Ramin},
  journal = {Phys. Rev. Lett.},
  volume = {99},
  issue = {4},
  pages = {048102},
  numpages = {4},
  year = {2007},
  month = {Jul},
  publisher = {American Physical Society},
  doi = {10.1103/PhysRevLett.99.048102},
}

@article{Vicsek1995,
  title = {Novel Type of Phase Transition in a System of Self-Driven Particles},
  author = {Vicsek, Tam\'as and Czir\'ok, Andr\'as and Ben-Jacob, Eshel and Cohen, Inon and Shochet, Ofer},
  journal = {Phys. Rev. Lett.},
  volume = {75},
  issue = {6},
  pages = {1226--1229},
  numpages = {0},
  year = {1995},
  month = {Aug},
  publisher = {American Physical Society},
  doi = {10.1103/PhysRevLett.75.1226},
}

@article{Wang2011spontaneous,
author = {Shenshen Wang  and Peter G. Wolynes },
title = {On the spontaneous collective motion of active matter},
journal = {Proceedings of the National Academy of Sciences},
volume = {108},
number = {37},
pages = {15184-15189},
year = {2011},
doi = {10.1073/pnas.1112034108}
}

@article{siwy2002origin,
  title={Origin of 1/f $\alpha$ noise in membrane channel currents},
  author={Siwy, Z and Fuli{\'n}ski, A},
  journal={Physical Review Letters},
  volume={89},
  number={15},
  pages={158101},
  year={2002},
  publisher={APS}
}

@article{Mognetti2013,
  title = {Living Clusters and Crystals from Low-Density Suspensions of Active Colloids},
  author = {Mognetti, B. M. and \ifmmode \check{S}\else \v{S}\fi{}ari\ifmmode \acute{c}\else \'{c}\fi{}, A. and Angioletti-Uberti, S. and Cacciuto, A. and Valeriani, C. and Frenkel, D.},
  journal = {Phys. Rev. Lett.},
  volume = {111},
  issue = {24},
  pages = {245702},
  numpages = {5},
  year = {2013},
  month = {Dec},
  publisher = {American Physical Society},
  doi = {10.1103/PhysRevLett.111.245702}
}

@article{Narayan2007,
author = {Vijay Narayan  and Sriram Ramaswamy  and Narayanan Menon },
title = {Long-Lived Giant Number Fluctuations in a Swarming Granular Nematic},
journal = {Science},
volume = {317},
number = {5834},
pages = {105-108},
year = {2007},
doi = {10.1126/science.1140414}
}

@article{Berthier2011,
  title={Overview of different characterisations of dynamic heterogeneity},
  author={Berthier, Ludovic and Biroli, Giulio and Bouchaud, Jean-Philippe and Jack, Robert L},
  journal={Dynamical heterogeneities in glasses, colloids, and granular media},
  volume={150},
  pages={68},
  year={2011},
  publisher={Oxford University Press Oxford}
}

@misc{SuppMat,
    note = {See Supplemental Material at [URL will be inserted by publisher] for analytical details, values of key experimental parameters and additional figures. The Supplemental Material includes Refs. \cite{Gomer1990,Hansen2013,Kawasaki1998,thorneywork2018structure}.}
}

@book{Hansen2013,
  title={Theory of simple liquids: with applications to soft matter},
  author={Hansen, Jean-Pierre and McDonald, Ian Ranald},
  year={2013},
  publisher={Academic press}
}

@article{Gomer1990,
  title={Diffusion of adsorbates on metal surfaces},
  author={R Gomer},
  journal={Reports on Progress in Physics},
  volume={53},
  number={7},
  pages={917--1002},
  year={1990}
}

@article{thorneywork2018structure,
	title = {Structure factors in a two-dimensional binary colloidal hard sphere system},
	volume = {116},
	url = {https://doi.org/10.1080/00268976.2018.1492745},
	doi = {10.1080/00268976.2018.1492745},
	number = {21-22},
	journal = {Molecular Physics},
	author = {Thorneywork, Alice L. and Schnyder, Simon K. and Aarts, Dirk G. A. L. and Horbach, Jürgen and Roth, Roland and Dullens, Roel P. A.},
	month = {nov},
	year = {2018},
	pages = {3245--3257},
}

@article{Kawasaki1998,
  title={Microscopic analyses of the dynamical density functional equation of dense fluids},
  author={Kawasaki, Kyozi},
  journal={Journal of statistical physics},
  volume={93},
  number={3},
  pages={527--546},
  year={1998},
  publisher={Springer}
}

@article{Kirchner2005,
  title={Aliasing in 1/ f $\alpha$ noise spectra: Origins, consequences, and remedies},
  author={Kirchner, James W},
  journal={Physical Review E—Statistical, Nonlinear, and Soft Matter Physics},
  volume={71},
  number={6},
  pages={066110},
  year={2005},
  publisher={APS}
}

@article{Mackay2024supp,
  title = {The Countoscope: Measuring Self and Collective Dynamics without Trajectories},
  author = {Mackay, Eleanor K. R. and Marbach, Sophie and Sprinkle, Brennan and Thorneywork, Alice L.},
  journal = {Phys. Rev. X},
  volume = {14},
  issue = {4},
  pages = {041016},
  numpages = {14},
  year = {2024},
  month = {Oct},
  publisher = {American Physical Society},
  doi = {10.1103/PhysRevX.14.041016},
  url = {https://link.aps.org/doi/10.1103/PhysRevX.14.041016}
}

@article{Dean1996,
  title={Langevin equation for the density of a system of interacting Langevin processes},
  author={Dean, David S},
  journal={Journal of Physics A: Mathematical and General},
  volume={29},
  number={24},
  pages={L613},
  year={1996},
  publisher={IOP Publishing}
}

@book{Bendat2011supp,
  title={Random data: analysis and measurement procedures},
  author={Bendat, Julius S and Piersol, Allan G},
  year={2011},
  publisher={John Wiley \& Sons}
}

\end{document}


\title{\textbf{Supplemental Material} \\  \vspace{2mm} Collective Hard Core Interactions Leave \\ Multiscale Signatures in Number Fluctuation Spectra} 

\author{Eleanor K. R. Mackay \and Anna Drummond Young \and Adam Carter \and Sophie Marbach \and Alice L. Thorneywork}
\date{}

\maketitle

\section{Link Between Correlation Function and Box Occupation Probability}
\label{sec:corr_to_prob}
To rationalise the shape of the PSD, we first note the link between the PSD and time autocorrelation function by the Wiener-Khinchin theorem, $S(f)=\int\langle N(t)N(0)\rangle e^{-i2\pi ft}\,dt$  \cite{Bendat2011supp}. The two scaling regimes in the PSD thus correspond to two distinct time scalings in the correlation function $\langle N(0)N(t)\rangle$. This can be recast in terms of individual particle contributions as $N(t)=\sum_in_i(t)$, where $n_i(t)=\{1,0\}$ when particle $i$ is inside or outside the box. Specifically, $n_i(t)=\int\int^{L/2}_{-L/2}\,\delta({\bf r}-{\bf r}_i)\,dxdy$. We expand the correlation function in order to separate two terms, due to single particle and particle-particle correlations in the total count:
\begin{equation*}
\begin{split}
    \langle N(0)N(t)\rangle  
     &= \sum_i \sum_j \langle n_i(0)n_j(t)\rangle \\
    &=\sum_{i=1}^\infty\langle n_i(t)n_i(0)\rangle + \sum_{i=1}^{\infty}\sum_{j\neq i}\langle n_j(t)n_i(0)\rangle \,\, ,
\end{split}
\end{equation*}

In a non-interacting system, different particles are uncorrelated and 
\begin{equation*}
    \langle N(t)N(0)\rangle=\sum_{i=1}^\infty\langle n_i(t)n_i(0)\rangle + \sum_{i=1}^{\infty}\sum_{j\neq i}\langle n_j(t)\rangle \langle n_i(0)\rangle =\sum_{i=1}^\infty\langle n_i(t)n_i(0)\rangle + N_L^2 ,
\end{equation*}
where $N_L$ is the mean number of particles in the box.

The expectation value $\langle n_i(0)n_i(t)\rangle$ can be evaluated as a weighted sum of its four possible values, where the weights are the probability of finding each value. Three of these possible values are zero, and so the expectation value reduces to the probability that the single particle count is $1$ at both $t=0$ and $t$, which we write as $P(n_i(0)\,\&\,n_i(t)=1)$. We have,
\begin{equation*}
\begin{split}
    \langle n_i(t)n_i(0)\rangle  
     &= P\left(n_i(t)=1 \, \& \, n_i(0)=1\right) \\
    &= P\left(n_i(t)=1 \, | \, n_i(0)=1\right)P\left(n_i(0)=1\right)
\end{split}
\end{equation*}
For the infinite steady state system, the probability $P\left(n_i(t)=1 \, | \, n_i(0)=1\right)$ is the same for all particles and time origins, and depends only on the time interval $t$. It is the integral of the single particle propagator over the box area, considering all possible start points within the box:
\begin{equation*}
\begin{split}
    P(n_i(t)=1 \, | \, n_i(0)=1) 
    &=  \frac{1}{L^2}\iiiint_{-L/2}^{L/2} \dd y \,\dd x \, \dd y' \, \dd x' \frac{1}{(\sqrt{4\pi Dt})^2} e^{- [(x-x')^2 + (y-y')^2]/4Dt}\\
    &=  \frac{1}{L^2}\left(\int_{-L/2}^{L/2} \dd x' \int_{-L/2 -x'}^{L/2-x'} \dd x  \frac{1}{\sqrt{4\pi Dt}} e^{- x^2/4Dt} \right)^2\\
    &= \left(\frac{1}{\sqrt{4\pi L^2 Dt}}\int_{-L/2}^{L/2}\int_{-L/2-x'}^{L/2-x'} \dd x \, \dd x' e^{-x^2/4Dt}\right)^2 \equiv P_{\mathrm{in}}(t) \,.
\end{split}
\end{equation*}
Finally,
\begin{equation}
    \langle N(t)N(0)\rangle - \meanN^2 = P_{\mathrm{in}}(t)\sum_{i=1}^\infty P(n_i(0)=1)=\meanN \, P_{\mathrm{in}}(t) \, ,
\end{equation}

where we have used
\begin{equation*}
    \begin{split}
        \meanN&\equiv \langle N(t)\rangle \\
        &= \sum_{i=1}^\infty \langle n_i(t) \rangle \\
        &= \sum_{i=1}^\infty \left[(1)P(n_i(t)=1) + (0)P(n_i(t)=0)\right] \\
        &= \sum_{i=1}^\infty P(n_i(t)=1)\, .
    \end{split}
\end{equation*}

In an interacting system, the term that senses particle-particle correlations can be evaluated in a similar way. We obtain
\begin{equation*}
\begin{split}
    \sum_{i=1}^\infty \sum_{j\neq i}^\infty \langle n_j(t)n_i(0)\rangle  
    &= \sum_{i=1}^\infty P\left(n_i(0)=1\right) \sum_{j\neq i}^\infty P\left(n_j(t)=1 \, | \, n_i(0)=1\right) \\
    &= \meanN \sum_{j\neq 1}^\infty P\left(n_j(t)=1 \, | \, n_1(0)=1\right) \\
    &\equiv N_L\Nnew \,,
\end{split}
\end{equation*}
where $n_1(t)$ is the indicator function for a single randomly selected particle. The total correlation function in an interacting system is therefore
\begin{equation}
    \langle N(t)N(0)\rangle =\meanN \, P_{\mathrm{in}}(t) + \meanN\Nnew \, .
\end{equation}

The functional form of $\Nnew$ is not known analytically; its behaviour is analysed experimentally in the main text and in Sec.~\ref{sec:supp:Pin_Nnew_phi}. Unlike $\Pin$, this term probes different transport processes depending on the box size. In small boxes ($L\lesssim\sigma$), $\Nnew$ is the total probability that exactly one particle that was not initially in the box now occupies the box at time $t$. Hard core interactions create this constraint, as only one particle can occupy the box at a time. $\Nnew$ in small boxes thus probes pairwise exchange. In contrast, multiple particles can contribute to $\Nnew$ in large boxes, which may or may not have initially occupied the box. It thus probes collective rearrangements in large boxes - specifically, the collective rearrangements around a single randomly selected particle. In this case, it ranges from $N_{\rm new}^{(\phi)}(t=0)=N_L-1$ at the initial time, because one random initial occupant is excluded from the particle count, to $\Nnew=N_L$, when the given initial particle has left the box.  This is the inverse of the tagged particle counts explored in Fig.~3 of the main text: $\Nnew$ is equivalent to a procedure where one random particle inside a box is tagged and systematically excluded from the particle count as a function of time.

\section{Analytical Methods}
\label{sec:theory}

We follow previous studies that calculate the correlation functions of number fluctuations in the framework of stochastic density field theory (sDFT)~\cite{Mackay2024supp,Carter2025,Minh2023}. A key advantage of the sDFT framework is that particle interactions can be included in Fourier space via the static structure factor $S(k)$ \cite{Hansen2013} (random phase approximation). The starting point is the Dean-Kawasaki equation~\cite{Dean1996,Kawasaki1998} for the time- and space-dependent particle number density $\rho(\mathbf{x},t)$ :
\begin{equation}
\label{eq:DK}
    \partial_t\rho(\mathbf{x},t) = D\nabla^2\rho(\mathbf{x},t) + \boldsymbol{\nabla}\cdot\left(\sqrt{2D\rho(\mathbf{x},t)}\boldsymbol{\xi}(\mathbf{x},t)\right) + 
    D\mathbf{\nabla}\cdot\left(\rho(\mathbf{x},t)\, \mathbf{\nabla}\int d\mathbf{x'}\rho(\mathbf{x'},t)\frac{U(\mathbf{x}-\mathbf{x'})}{k_BT}\right)
\end{equation}
where $D$ is the self diffusion coefficient, $\boldsymbol{\xi}$ is a Gaussian white noise force and $U(\mathbf{r})$ is the pair interaction potential. Density fluctuations $\psi(\mathbf{x},t) = \rho(\mathbf{x},t) - \rho_0$, where $\rho_0$ is the mean density, can be integrated over the domain of interest to characterise the number fluctuations. Indeed, in 2D, 
\begin{equation}
    N(t) = \iint_{-L/2}^{L/2} \rho(\mathbf{x},t) \dd \mathbf{x} = \iint_{-L/2}^{L/2} \left(\rho_0 + \psi(\mathbf{x},t)\right) \dd \mathbf{x} \equiv \meanN + \iint_{-L/2}^{L/2} \psi(\mathbf{x},t) \dd \mathbf{x},
\end{equation}
where $L^2 \rho_0\equiv\meanN$ is the mean number of particles in a box. 
We can therefore study the fluctuations and correlations of $\psi(\mathbf{x},t)$ to obtain that of $N(t)$, since
\begin{equation}
\label{eq:ncorr}
    \langle N(t) N(t') \rangle = \meanN^2 + \iiiint_{-L/2}^{L/2} \dd\mathbf{x} \dd\mathbf{x}' \langle \psi(\mathbf{x},t) \psi(\mathbf{x}',t') \rangle. 
\end{equation}
Note that we write for simplicity analytical derivations in a square box, but that these derivations are straightforwardly extended to rectangular boxes of shape $L_1\times L_2$.

\subsection{Linear Order Density Fluctuations}
We consider density fluctuations $\psi(\mathbf{x},t) = \rho(\mathbf{x},t) - \rho_0$ around a uniform, constant mean density. Expanding Eq.~\eqref{eq:DK} to linear order in $\psi$:
\begin{equation*}
    \partial_t\psi(\mathbf{x},t) = D\nabla^2\psi(\mathbf{x},t) + \boldsymbol{\nabla}\cdot\left(\sqrt{2D\rho_0}\,\boldsymbol{\xi}(\mathbf{x},t)\right) + 
    D \rho_0\mathbf{\nabla}^2\int d\mathbf{x'}\psi(\mathbf{x'},t)\frac{U(\mathbf{x}-\mathbf{x'})}{k_BT}
\end{equation*}
and taking the Fourier transform:
\begin{equation}
    i\omega\,\widetilde{\psi}=-Dk^2\widetilde{\psi}+\sum_{j}ik_j\sqrt{2D\rho_0}\,\widetilde{\xi}_j - D\rho_0k^2\frac{\widetilde{U}}{k_BT}\widetilde{\psi} .
\end{equation}

We use the white noise characteristics $\langle\xi_j\xi'_k\rangle=\delta_{jk}\delta(\omega+\omega')\delta^d(\mathbf{k}+\mathbf{k'})$ to evaluate the correlation function. The random phase approximation $S(k)=1\,/\left(1+\frac{\rho_0\widetilde{U}}{k_BT}\right)$ lets us link the particle interactions to the static structure factor. The result is:
\begin{equation}
\label{eq:psicorr}
    \langle\widetilde{\psi}(\mathbf{k},\omega)\widetilde{\psi}(\mathbf{k'},\omega')\rangle=2\rho_0\frac{Dk^2}{\omega^2+(\frac{Dk^2}{S(k)})^2}\delta^d(\mathbf{k}+\mathbf{k'})\delta(\omega+\omega').
\end{equation}

\subsection{Correlation Function for Number Correlations}
\label{sec:CN}

Using Eq.~\ref{eq:psicorr} and inserting in Eq.~\ref{eq:ncorr}, the correlation function for numbers is obtained as
\begin{equation}
    \langle N(t)N(t')\rangle-\meanN^2=\meanN\int\frac{k \dd k}{(2\pi)^2}f_\mathscr{V}(k)S(k)e^{-Dk^2|t-t'|/S(k)},
    \label{eq:Ngeneral}
\end{equation}
with
\begin{equation*}
    f_\mathscr{V}(k)=L^2\int d\theta  \left(\frac{2\mathrm{sin}(k \cos \theta L/2)}{k\cos \theta L}\right)^2\left(\frac{2\mathrm{sin}(k \sin \theta L/2)}{k\sin \theta L}\right)^2,
\end{equation*}
where we recall that $S(k)$ is the static structure factor.

In the dilute regime, $S(k) \simeq 1$ and it can be shown~\cite{Mackay2024supp} that the correlation function satisfies
\begin{equation}
    C_N(t) = \langle N(t) N(0) \rangle - \meanN^2 = f(4Dt/L_1^2) f(4Dt/L_2^2)
\end{equation}
in the case where the box is of size $L_1\times L_2$ and
where the function $f$ is
\begin{equation}
     f\left(\tau = \frac{4 D t}{L^2}\right) = \sqrt{\frac{\tau}{\pi}} \left( e^{-1/\tau} - 1\right) + \mathrm{erf} \left( \sqrt{1/\tau} \right).
     \label{eq:famousfunction}
\end{equation}
One can then show that in the long time limit, $f(\tau\rightarrow\infty) \simeq 1/\sqrt{\pi \tau}$, such that the correlation function for a square box at long times satisfies
\begin{equation}
    C_N(t) \underset{t\rightarrow\infty}{=} \frac{L^2}{4\pi D t} + o\left(\frac{1}{t}\right).
\end{equation}
In contrast, the short time expansion yields
\begin{equation}
    C_N(t) \underset{t\rightarrow 0}{=} 1 - \frac{8}{\sqrt{\pi}}\meanN \sqrt{\frac{Dt}{L^2}} + O\left(t\right).
\end{equation}

\subsection{Static Structure Factor}
\label{sec:supp:Sk_phi}

In Eq.~\eqref{eq:psicorr} and onwards we use an analytic expression for the structure factor of hard spheres in 2D, which is based on density field theory and is in remarkable agreement with our 2D-sedimented colloidal experiments, as was verified in previous works~\cite{thorneywork2018structure,Carter2025}. We report it here for consistency:
\begin{equation}
    S(k) = \frac{1}{1 - \rho c^{(2)}(k)}
    \label{eq:S of k}
\end{equation}
where $\rho$ is the mean particle density and 
\begin{equation}
\begin{split}
        c^{(2)}(k) = &\frac{\pi}{6(1-\phi)^3 k^2} \bigg[ - \frac{5}{4} (1-\phi)^2 k^2 \sigma^2 J_0(k\sigma/2)^2 \\
        &+ \bigg( 4 \left( \,( \phi - 20) \phi + 7 \right) \\
        & + \frac{5}{4}(1 - \phi)^2 k^2 \sigma^2 \bigg) J_1(k\sigma/2)^2 \\
        &+ 2(\phi - 13)(1 - \phi) k \sigma J_1(k\sigma/2) J_0 (k\sigma/2) \bigg]
\label{eq:Sk}
\end{split}
\end{equation}
where $J_i(x)$ are Bessel functions of the first kind. The limit of vanishing wavenumber in Eq.~\eqref{eq:Sk} can be taken analytically
\begin{equation}
    S(k = 0) = \frac{(1- \phi)^3}{1 + \phi}
    \label{eq:sk0}
\end{equation}
and is consistent with the result from the scaled particle theory equation of state.

\subsection{Power Spectral Density of $N(t)$}
In this work we consider the Power Spectral Density of the particle count $N(t)$, the N-PSD. This is equal to the Fourier transform of the autocorrelation function,
\begin{equation*}
    S(\omega) = \int_{-\infty}^{\infty}\langle N(t)N(0)\rangle e^{-i\omega t}\,dt.
\end{equation*}
Constant terms contribute only delta functions $\delta(\omega)$ to the N-PSD, and we leave them out of the following analysis. Using Eq.~\eqref{eq:Ngeneral}, and taking its temporal Fourier transform, one directly arrives at 

\begin{equation}
\label{eq:SNf}
    S(\omega)=\meanN \int\frac{ k \,\dd k}{(2\pi)^2}f_{\mathcal{V}}(k)\frac{2 Dk^2}{\omega^2+(\frac{Dk^2}{S(k)})^2}.
\end{equation}

\subsection{Low Frequency and High Frequency Limits Excluding Interactions}

We can obtain some of the limiting behaviour of the N-PSD from the expression in Eq.~\eqref{eq:SNf}, in 1D in a case with no interactions. 
We rewrite the 1D equivalent of Eq.~\ref{eq:SNf} as:
\begin{equation*}
    S(\omega)=2\rho_0\sqrt{D}\omega^{-3/2}\int\frac{\dd u}{2\pi}\frac{u^2}{1+u^4}\left(\frac{2\mathrm{sin}\left(\sqrt{\frac{L^2\omega}{4D}}\,u\right)}{u}\right)^2,
\end{equation*}
where we defined $u = \sqrt{D/\omega}\,k$ and we have taken $S(k)=1$.

For $\omega\gg\frac{4D}{L^2}$, \textit{i.e.} the high frequency limit, we can approximate $\mathrm{sin}^2\left(\sqrt{\frac{L^2\omega}{4D}}u\right)\approx 1/2$:
\begin{equation*}
    S(\omega)\approx2\rho_0\sqrt{D}\omega^{-3/2}\int_{-\infty}^{\infty}\frac{\dd u}{\pi}\frac{1}{1+u^4}=\frac{\sqrt{2}\meanN L^2}{D}\left(\frac{L^2\omega}{D}\right)^{-3/2}
\end{equation*}
and we indeed recover the short time, high frequency scaling as $1/\omega^{3/2}$.

For $\omega\ll\frac{4D}{L^2}$, \textit{i.e.} the low frequency limit, we can approximate $\mathrm{sin}^2\left(\sqrt{\frac{L^2\omega}{4D}}\,u\right)\approx \frac{L^2\omega}{4D}u^2$:
\begin{equation*}S(\omega)\approx4\rho_0\sqrt{D}\omega^{-3/2}\frac{L^2\omega}{4D}\int_{-\infty}^{\infty}\frac{\dd u}{\pi}\frac{u^2}{1+u^4}=\frac{\meanN L^2}{\sqrt{2}D}\left(\frac{L^2\omega}{D}\right)^{-1/2}.
\end{equation*}
Here we recovered the $1/\omega^{1/2}$ scaling expected at low frequencies in 1D. Note we can measure these regimes in Fig. 2(b) in the main paper. The integral is harder to calculate in two dimensions. Instead, we infer the low and high frequency scalings from the long and short time scalings of the correlation function, see Section~\ref{sec:CN}.

\section{Experimental analysis}
\subsection{Calculation of the PSD}

Power spectral densities are calculated using Welch's method \cite{Press1992} as implemented in the open source python package scipy. Fundamentally, this calculates the PSD from the complex conjugate of the Fourier transform of $N(t)$. The method uses a window function to minimise spurious high frequency scalloping associated with the signal's finite duration, for which we choose the periodic Hann window. The method also averages the PSD for multiple overlapping section of the signal: we use a segment length of $2^{16}=65536$ data points and an overlap of 32768 data points.

Raw power spectral densities are then filtered with anti-aliasing filters, similar to \cite{Kirchner2005}. These remove a `tail' at high frequencies, where noise for frequencies higher than the sampling frequency, $f_s=2\,{\rm fps}$ is aliased back into the measured frequency range. The resulting PSDs are further smoothed by logarithmically binning along the frequency axis and taking the average noise power in each bin.

Finally, to estimate the PSD at the very lowest frequencies, those set by the total duration of the experiment, we directly calculate the discrete Fourier transform of measured $N(t)$ as
\begin{equation}
    \mathscr{F}(N)=\sum_{m=0}^M N(t_m) \Delta t \exp\left( -\frac{2\pi i \,km}M\right)
\end{equation}

where $\Delta t$ is the time between samples and $M=T/\Delta t$ is the total number of samples. The frequency is given by $f=k/T$ and we calculate for $k=\{1,2,3,4,5\}$. The PSD at these frequencies is then given by $S(f)=\mathscr{FF^*}/T$.

\subsection{Statistical sampling}

To probe the full range of spectral scalings in the N-PSD, we consider length scales $L=0.3-50\,{\rm \mu m}$. For each box size, we consider large numbers of boxes placed at different locations within an experimental image, which is $370\,{\rm \mu m} \times 295\,{\rm \mu m}$ in our experiments. By tiling large numbers of boxes on to each image, we significantly improve the statistical resolution \cite{Carter2025}. We use $\sim 10^4$ boxes for the smallest box sizes and $\sim 10^3$ boxes for the largest box sizes, where the areas of these large boxes must partially overlap in order to fit within the image \cite{Carter2025}.

Images are recorded at the centre of a macroscopic colloidal monolayer, $38\,{\rm mm}\times 9\,{\rm mm}$. As the edges of the system are $\sim 10^3$ particle diameters away from the imaged region, we can neglect the finite size of the monolayer.

\subsection{Experimental Diffusion Coefficients}

\begin{table}[h!]
    \centering
    \begin{tabular}{c | c | c | c}
        $\phi$ & $D$ / ${\rm \mu m^2s^{-1}}$ & $D_l$ / ${\rm \mu m^2s^{-1}}$ & $D_{\rm coll}$ / ${\rm \mu m^2s^{-1}}$ \\
         \hline
         0.02 & 0.049 (0.003) & / & / \\
         0.07 & 0.043 (0.003) & 0.034 (0.003) & 0.06 (0.009) \\
         0.12 & 0.043 (0.001) & 0.032 (0.001) & 0.07 (0.01) \\
         0.39 & 0.030 (0.002) & 0.020 (0.001) & 0.19 (0.03) \\
         0.60 & 0.022 (0.002) & 0.009 (0.0006) & 0.23 (0.04) \\
    \end{tabular}
    \caption{Diffusion coefficients in experimental systems. 
    Errors are given in brackets, and are estimated from a combination of the statistical error on fitting to data and the uncertainty in calibrating the $\mu$m:pixel conversion for our microscope (1\%, which is an improvement on our previous work in Ref.~\cite{Mackay2024supp} due to improvements in methodology).}
    \label{tab:D values}
\end{table}
The short-time self ($D$), long-time self ($D_l$) and collective diffusion coefficients ($D_{\rm coll}$) are reported in Table~1 for each experimental system. Self diffusion coefficients are extracted at short times by a linear fit to the MSD and at long times as the limit $\lim_{t\rightarrow\infty}\langle r^2\rangle/(4t)$. Collective diffusion coefficients are calculated as $D\,(1+\phi)/(1-\phi)^3$\cite{Carter2025}.

\section{Further details of experimental spectra}
\subsection{Trends in Small Box $P_{\rm in}(t)$ and $N_{\rm new}^{(\phi)}(t)$ at Varying Packing Fraction}
\label{sec:supp:Pin_Nnew_phi}

\begin{figure}
    \centering
    \includegraphics[width=\linewidth]{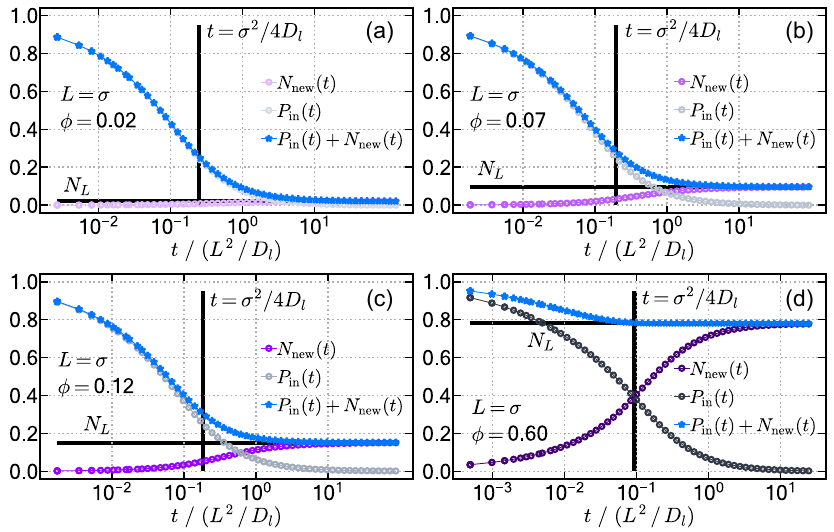}
    \caption{{\bf Long time scaling at small length scales depends on packing fraction.} Single particle ($\Pin$, grey points), interparticle ($\Nnew$, purple points) and total ($\Pin + \Nnew$, blue stars) correlations in experimental systems at a range of $\phi$ for $L=\,\sigma$. Shades of grey and purple correspond to packing fraction. Black lines indicate the time scale $\sigma^2/(4D_l)$, where $D_l$ are the experimental long-time diffusion coefficients.}
    \label{fig:Nnew_phi}
\end{figure}


In Fig.~\ref{fig:Nnew_phi} we show $\Pin$ and $\Nnew$ in boxes of length $L=\sigma$ for the experimental systems at $\phi = [0.02, 0.08, 0.12, 0.60]$. These two contributions to the total correlation function again approach the same long time behaviour for $\phi = 0.6$ (appearing as reflections of each other on a linear scale), creating a regime in which the correlation function rapidly approaches $\meanN$, as for $\phi=0.39$. At lower packing fractions, however, the magnitude of the $1/t$ decay at long times is smaller in $\Nnew$ than $\Pin$. This means that the total correlation function continues to show a $1/t$ time dependence at long times, as correlations in $\Pin$ are not suppressed by anti-correlations in $\Nnew$.

\subsection{Spectral Scalings in Large Boxes at Varying Packing Fraction}

\begin{figure}[h]
    \centering
    \includegraphics[width=0.85\linewidth]{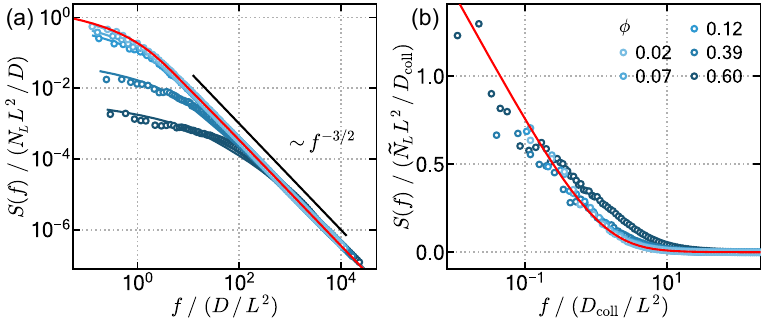}
    \caption{\textbf{High- and low-frequency spectral scalings for $L\gg \sigma$.} (a) Power spectral densities of particle number $N(t)$ in boxes $L=7.6\,\sigma$ for $\phi=0.02-0.60$, where colour indicates packing fraction. Amplitude and frequency have been rescaled, where $\meanN$ is the mean number of particles, $L$ is the box size and $D$ is the experimental short-time self-diffusion coefficient. Points show experimental data and solid lines show theoretical results from sDFT including hard disk interactions. Red solid line shows theoretical results from sDFT without interactions. (b) Same data on linear-logarithmic axes, now rescaled using $\tilde{N}_L = \meanN(1-\phi)^3/(1+\phi)$ and $D_{\rm coll}=(1+\phi)/(1-\phi)^3$ \cite{Carter2025}. Red solid line shows theoretical predictions from sDFT without interactions.}
    \label{fig:supp:spectra_large}
\end{figure}

In the main text, we analyse the mechanisms that govern the spectral scalings for boxes $L \lesssim \sigma$ as a function of $\phi$. Here, we show the corresponding spectral scalings for large boxes, $L=7.6\,\sigma$, and directly compare the spectra for multiple packing fractions, $\phi=0.02-0.60$.
In Fig.~\ref{fig:supp:spectra_large}(a), we show spectra rescaled with a characteristic frequency $f_D=D/L^2$ and noise amplitude $N_L / f_D$. This rescaling collapses all data to the same curve at high frequencies, where we see a characteristic $S(f)\sim f^{-3/2}$ scaling. Fluctuations in $N(t)$ over these time scales are mediated by particles diffusing to the edge of the box \cite{Marbach2021,Minh2023,Mackay2024supp}. At low frequencies, however, curves show similar scaling behaviour, but do not collapse. This is a consequence of the fact that, at low frequencies, behaviour is governed by the collective diffusion coefficient, and this has a different functional dependence on $\phi$ to the self-diffusion coefficient. 

In Fig.~\ref{fig:supp:spectra_large}(b) we show the same spectra, now on linear-logarithmic axes. This highlights the $\sim -\log(f)$ scaling at low frequencies.
Data are rescaled by a characteristic frequency $f_{\rm coll}=D_{\rm coll}/L^2$ and characteristic noise amplitude $\tilde{N}_L/f_{\rm coll}$, which collapses the data except for at the highest value of $\phi$. Here, $\tilde{N}_L$ is the variance in particle number and, in large boxes, is given by $\tilde{N}_L=N_L\,S(k\rightarrow 0)$  \cite{thorneywork2018structure} and the collective diffusion coefficient is determined as $D_{\rm coll}=D/S(k\rightarrow 0)$ \cite{Gomer1990,Carter2025}. In both cases, we use a theoretical prediction for $S(k\rightarrow 0)$ for an interacting hard disk system, $S(k \rightarrow 0)=(1-\phi)^3/(1+\phi)$, see Sec.~\ref{sec:supp:Sk_phi}.

Physically, the fact that we recover the $S(f)\sim -\log(f)$ scaling in this frequency regime means we have accessed the time and length scales over which the many-body system is effectively a continuum fluid, described with a number density $\rho$, where fluctuations in this number density relax diffusively \cite{Dhont1996,Carter2025}. These are length scales longer than the range of the interaction between the Brownian particles \cite{Dhont1996} and time scales long enough for particles to interact directly with their neighbours. 

\subsection{Time Correlation Functions}

\begin{figure}
    \centering
    \includegraphics[width=\linewidth]{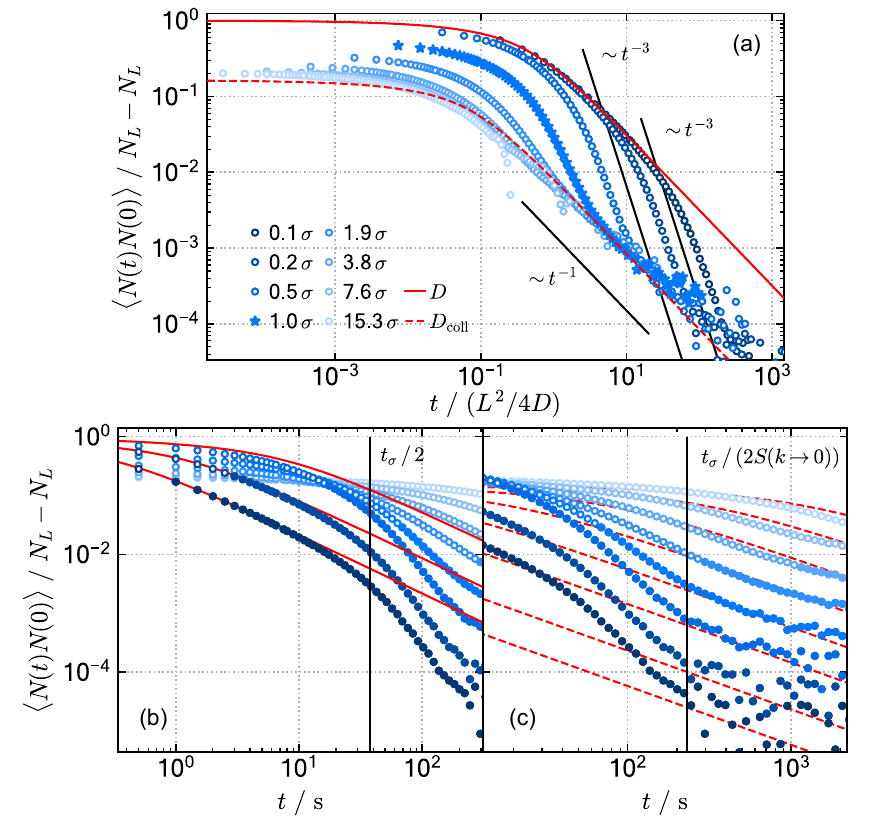}
    \caption{\textbf{Time correlation functions sense self-diffusion at short time scales and collective diffusion at long time scales} Experimental data for $\phi=0.39$. (a) Points show experimental correlation functions calculated for a range of box sizes. Colour corresponds to box size, as given in the legend. Correlation functions are plotted as $\langle N(t)N(0)\rangle/\meanN - \meanN$ to show the long-time power law scaling. Time is rescaled by $t_D=L^2/(4D)$. Red lines indicate the correlation function for analytical theory in the absence of interactions. The solid line takes the short-time self diffusion coefficient, $D$, while the dashed line takes the collective diffusion coefficient for hard disk interactions, $D_{\rm coll}=D(1+\phi)/(1-\phi)^3$ \cite{Gomer1990,thorneywork2018structure}. (b) Data as in (a) with unscaled time axis in the limit of short times. Points are filled in for times $t>L^2/(4D)$. Red lines show non-interacting theory taking the short-time self-diffusion coefficient, $D$, as input. Black line indicates the time scale $t_\sigma /2=\sigma^2/(4D)/2$. (c) Data as in (a) with unscaled time axis in the limit of long times. Red dashed lines show non-interacting theory taking the collective diffusion coefficient, $D_{\rm coll}$, as input. Black line indicates $t_\sigma/S(k\rightarrow 0)/2$. }
    \label{fig:supp:real_t_corr}
\end{figure}

The power spectral density of a fluctuating signal is related to the time autocorrelation function by a Fourier transform \cite{Bendat2011supp}. 
While time- and frequency-space quantities in principle contain the same information, different correlation functions naturally highlight  different features. 
Here, for reference, we show in Fig.~\ref{fig:supp:real_t_corr} details of the corresponding time correlation functions $\langle N(t)N(0)\rangle$ for the experimental system at $\phi=0.39$.

At long times, the correlation function $\langle N(t)N(0)\rangle$ decays towards a finite value $\langle N\rangle^2=N_L^2$. To resolve relevant temporal scalings, we thus plot the rate at which the correlation function approaches this value, as shown in Fig.~\ref{fig:supp:real_t_corr}(a).
At short times, the time-correlation function $\langle N(t)N(0)\rangle$ decays from a finite value $\langle N^2\rangle$. This masks the relevant short-time $t^{1/2}$ power law scaling regime that appears as an $1/f^{3/2}$ scaling at high frequencies in the spectra. Short-time scalings are better observed in the time domain via the quantity $\langle\Delta N^2(t)\rangle$, as explored in previous work \cite{Mackay2024supp}. 
The characteristic time scale corresponding to the short time regime is $L^2/(4D)$, which is shown by the analytical theory (red solid line) for a non-interacting system using the short-time self-diffusion coefficient as input. This agrees well with the time correlation function for small boxes at short times. 
For large boxes $L\gg \sigma$ at long times, the correlation functions all tend to the dashed red curve corresponding to collective diffusion. This is consistent with the fact that the low frequency regime of the PSDs for large boxes collapse on rescaling by a characteristic frequency $f_{\rm coll}=D_{\rm coll}/L^2$ and a characteristic noise amplitude $\tilde{N}_L/f_{\rm coll}$, as in Fig.~\ref{fig:supp:spectra_large}(a).

The long-time behaviour is more complex in small boxes $L\lesssim\sigma$. At intermediate times, the correlation functions decay rapidly, approaching $\meanN^2$ like $1/t^3$. This regime corresponds to the pairwise exchange kinetics discussed in the main text, which produces a plateau in the N-PSD. At long times, the small box correlation functions finally join the collective curve, in line with the behaviour in large boxes. While in principle a $-\log f$ scaling would therefore emerge in small-box N-PSDs at low frequencies, this is outside the range of frequencies probed by our experiment: the N-PSD, being an integrated quantity over time, is dominated by the rapidly decaying regime at intermediate times, which creates the finite plateau in frequency space.

To better pinpoint the dominant time scales underlying these regimes, Fig.~\ref{fig:supp:real_t_corr}(b) and (c) show the short and long time regime of the time correlation function with unscaled time axis. In panel (b), where we compare the correlation functions to the non-interacting theory, the initial regime governed by self-diffusion persists up to $t\sim t_\sigma=\sigma^2/(4D)$ for small boxes. As discussed in the main text, this is the time scale at which new particles are first able to enter the box, and where $\Nnew$ therefore becomes significant. In contrast, in panel (c) we compare the correlation functions to the expected theoretical dependence for behaviour governed by collective dynamics. Here, we see that curves appear to meet the theoretical curves at a time scale $t\sim t_\sigma/S(k\rightarrow 0)$. This time scale corresponds to $t\sim \xi^2/D_{\rm coll}$, where the length scale $\xi \sim \sigma/S(k\rightarrow 0)$ has been found to describe the size of transiently ordered regions in hard disk fluids \cite{Mackay2024supp}. The relaxation of these transiently ordered regions is mediated by collective rearrangements, and the time $t\sim \xi^2/D_{\rm coll}$ can therefore be rationalised as the time for relaxation of these transient ordered structures via collective diffusion.

\typeout{}
\bibliographystyle{plain}
\bibliography{suppbib}